\documentclass[11pt,onecolumn]{IEEEtran}
\usepackage{amsthm}
\usepackage{times,amssymb,amsmath,amsfonts,float,nicefrac,color,bbm,mathrsfs,caption,float}
\usepackage{algorithm,enumerate,multirow,caption,tikz,graphicx}
\usepackage[mathscr]{eucal}
\usepackage{sidecap}
\usepackage{algpseudocode}
\usepackage{verbatim}
\usepackage{epstopdf}
\usepackage{textcomp}
\usetikzlibrary{shapes,arrows}
\usepackage{stmaryrd}
\usepackage{mathabx}
\usepackage[noadjust]{cite}
\usepackage{booktabs}
\usepackage[normalem]{ulem}
\usepackage[top=1in,bottom=1in,left=1in,right=1in]{geometry}
\allowdisplaybreaks

\newtheorem{theorem}{Theorem}[section]

\newtheorem{lemma}[theorem]{Lemma}
\newtheorem{proposition}[theorem]{Proposition}
\newtheorem{definition}[theorem]{Definition}

\newtheorem{remark}[theorem]{Remark}

\newcommand{\RN}[1]{%
  \textup{\expandafter{\romannumeral#1}}%
}

\newcommand\remove[1]{}
\newcommand{\nc}{\newcommand}

\allowdisplaybreaks
\nc\bfa{{\boldsymbol a}}\nc\bfA{{\boldsymbol A}}\nc\cA{{\mathcal A}}\nc\sA{{\mathscr A}}
\nc\bfb{{\boldsymbol b}}\nc\bfB{{\boldsymbol B}}\nc\cB{{\mathcal B}}\nc\sB{{\mathscr B}}
\nc\bfc{{\boldsymbol c}}\nc\bfC{{\boldsymbol C}}\nc\cC{{\mathcal C}}\nc\sC{{\mathscr C}}
\nc\bfd{{\boldsymbol d}}\nc\bfD{{\boldsymbol D}}\nc\cD{{\mathcal D}}
\nc\bfe{{\boldsymbol e}}\nc\bfE{{\boldsymbol E}}\nc\cE{{\mathcal E}}
\nc\bff{{\boldsymbol f}}\nc\bfF{{\boldsymbol F}}\nc\cF{{\mathcal F}}\nc\sF{{\mathscr F}}
\nc\bfg{{\boldsymbol g}}\nc\bfG{{\boldsymbol G}}\nc\cG{{\mathcal G}}
\nc\bfh{{\boldsymbol h}}\nc\bfH{{\boldsymbol H}}\nc\cH{{\mathcal H}}
\nc\bfi{{\boldsymbol i}}\nc\bfI{{\boldsymbol I}}\nc\cI{{\mathcal I}}\nc\sI{{\mathscr I}}
\nc\bfj{{\boldsymbol j}}\nc\bfJ{{\boldsymbol J}}\nc\cJ{{\mathcal J}}
\nc\bfk{{\boldsymbol k}}\nc\bfK{{\boldsymbol K}}\nc\cK{{\mathcal K}}
\nc\bfl{{\boldsymbol l}}\nc\bfL{{\boldsymbol L}}\nc\cL{{\mathcal L}}
\nc\bfm{{\boldsymbol m}}\nc\bfM{{\boldsymbol M}}\nc\cM{{\mathcal M}}
\nc\bfn{{\boldsymbol n}}\nc\bfN{{\boldsymbol N}}\nc\cN{{\mathcal N}}
\nc\bfo{{\boldsymbol o}}\nc\bfO{{\boldsymbol O}}\nc\cO{{\mathcal O}}
\nc\bfp{{\boldsymbol p}}\nc\bfP{{\boldsymbol P}}\nc\cP{{\mathcal P}}\nc\eP{{\EuScriptP}}\nc\fP{{\mathfrak P}}
\nc\bfq{{\boldsymbol q}}\nc\bfQ{{\boldsymbol Q}}\nc\cQ{{\mathcal Q}}
\nc\bfr{{\boldsymbol r}}\nc\bfR{{\boldsymbol R}}\nc\cR{{\mathcal R}}\nc\sR{{\mathscr R}}
\nc\bfs{{\boldsymbol s}}\nc\bfS{{\boldsymbol S}}\nc\cS{{\mathcal S}}
\nc\bft{{\boldsymbol t}}\nc\bfT{{\boldsymbol T}}\nc\cT{{\mathcal T}}
\nc\bfu{{\boldsymbol u}}\nc\bfU{{\boldsymbol U}}\nc\cU{{\mathcal U}}
\nc\bfv{{\boldsymbol v}}\nc\bfV{{\boldsymbol V}}\nc\cV{{\mathcal V}}\nc\sV{{\mathscr V}}
\nc\bfw{{\boldsymbol w}}\nc\bfW{{\boldsymbol W}}\nc\cW{{\mathcal W}}\nc\sW{{\mathscr W}}
\nc\bfx{{\boldsymbol x}}\nc\bfX{{\boldsymbol X}}\nc\cX{{\mathcal X}}
\nc\bfy{{\boldsymbol y}}\nc\bfY{{\boldsymbol Y}}\nc\cY{{\mathcal Y}}
\nc\bfz{{\boldsymbol z}}\nc\bfZ{{\boldsymbol Z}}\nc\cZ{{\mathcal Z}}

\DeclareMathOperator{\tr}{tr}

\DeclareMathOperator{\wt}{wt}

\begin{document}


\title{Error correction based on partial information}

\author{\IEEEauthorblockN{Itzhak Tamo} \hspace*{1in}
\and \IEEEauthorblockN{Min Ye} \hspace*{1in}
\and \IEEEauthorblockN{Alexander Barg}}

\maketitle
{\renewcommand{\thefootnote}{}\footnotetext{

\vspace{-.2in}
 
\noindent\rule{1.5in}{.4pt}

A preliminary version of this paper was presented at the 2017 IEEE International Symposium on Information Theory, June 2017, Aachen, Germany \cite{Tamo17}.

I. Tamo is with Department of EE-Systems, Tel Aviv University, Tel Aviv,
Israel. Email: zactamo@gmail.com.
His research is supported by ISF grant no.~1030/15 and the NSF-BSF grant no.~2015814.

M. Ye was with Department of ECE and ISR, University of Maryland, College Park, MD 20742. Email: yeemmi@gmail.com
His research was supported by NSF grant CCF1422955.

A. Barg is with Department of ECE and ISR, University of Maryland, College Park, MD 20742 and also with IITP, Russian Academy of Sciences, 127051 Moscow, Russia. Email: abarg@umd.edu. His research was supported by NSF grants CCF1814487, CCF1618603, and CCF1422955.}
}
\renewcommand{\thefootnote}{\arabic{footnote}}
\setcounter{footnote}{0}

\begin{abstract}
We consider the decoding of linear and array codes from errors when we are only allowed to download a part of the codeword. More specifically, suppose that we have encoded $k$ data symbols using an $(n,k)$  code with code length $n$ and dimension $k.$ During storage, some of the codeword coordinates might be corrupted by errors. We aim to recover the original data by reading the corrupted codeword with a limit on the transmitting bandwidth, namely, we can only download an $\alpha$ proportion of the corrupted codeword. For a given $\alpha,$ our objective is to design a code and a decoding scheme such that we can recover the original data from the largest possible number of errors. A naive scheme is to read $\alpha n$ coordinates of the codeword. This method used in conjunction with MDS codes guarantees recovery from any $\lfloor(\alpha n-k)/2\rfloor$ errors. In this paper we show that we can instead read an $\alpha$ proportion from each of the codeword's coordinates. For a well-designed MDS code, this method can guarantee recovery from $\lfloor (n-k/\alpha)/2 \rfloor$ errors, which is $1/\alpha$ times more than the naive method, and is also the 
maximum number of errors that an $(n,k)$ code can correct by downloading only an $\alpha$ proportion of the codeword. We present two families of such optimal constructions and decoding schemes. One is a  Reed-Solomon code with evaluation points in a subfield and the other is based on Folded Reed-Solomon codes. We further show that both code constructions attain asymptotically optimal list decoding radius when downloading only a part of the corrupted codeword. We also construct an ensemble of random codes that with high probability approaches the upper bound on the number of correctable errors when the decoder downloads an $\alpha$ proportion of the corrupted codeword.
\end{abstract}

\section{Introduction}
Recovery of information under limitations on the repair bandwidth has received signification attention in information
theory literature. In particular, a well-known approach to enhance resilience of distributed storage systems against
failures of storage disks relies on Maximum Distance Separable (MDS) codes which are optimal in terms of the redundancy-reliability tradeoff.
 More specifically, an MDS code with $r$ parity symbols can recover the original data from any $r$ erasures of the codeword coordinates. In practice, single disk failure is the most common scenario. Upon observing this, {\sc Dimakis} et al. \cite{Dimakis10} introduced the concept of repair bandwidth, which is the minimum possible amount of data one needs to download in order to recover any single node failure. An MDS codes with optimal (minimum) repair bandwidth is called Minimum Storage Regenerating (MSR) code. In the low rate regime, {\sc Rashmi} et al. gave an explicit construction of MSR codes \cite{Rashmi11}. Constructions optimal-repair regenerating codes with no limitations on the code rate were given in several works of the authors \cite{Tamo13, Ye16, Ye16b,Ye18}. {\sc Guruswami} and {\sc Wootters} studied the repair bandwidth of Reed-Solomon (RS) codes \cite{Guruswami15}. Constructions of RS codes with optimal repair bandwidth were given in \cite{Tamo17RS,Ye16c,Tamo18}.

In this paper we consider the problem of decoding linear and array codes from errors when we are allowed to rely only on a part of the corrupted codeword. If we encode the original data using an $(n,k)$ MDS code with code length $n$ and dimension $k,$ it is well known that we can recover the original data from any $\lfloor (n-k)/2 \rfloor$ errors when we receive the whole codeword. In a distributed system, reading the whole codeword requires certain amount of disk I/Os and transmitting bandwidth. Now suppose that we have a limit on the bandwidth and we can only download an $\alpha<1$ proportion of the whole codeword, a natural question is how many errors we can guarantee to correct in this setup. In other words, how much error correcting capability is sacrificed by reducing the transmitting bandwidth.

Similarly to the study of MSR codes, we also resort to {\em array codes} \cite{Blaum98}.
An $(n,k,l)$ array code $\cC$ over a finite field $F$ is formed of $l\times n$ matrices $(C_1,\dots,C_n)\in (F^l)^n$. Each column $C_i$ of the matrix is a codeword coordinate, and the parameter $l$ that determines the dimension of the column vector $C_i$ is called {\em sub-packetization}.
Note that a scalar code can also be viewed as an array code with $l=1.$

\begin{definition}[Fractional decoding and $\alpha$-decoding radius]\label{def:radius} 
Consider an $(n,k,l)$ array code $\cC=\{(C_1,\dots,C_n)\}$ over $F$, 
where $C_i\in F^l, i=1,\dots, n$. 

(i) We say that $\cC$ can correct up to $t$ errors by downloading an $\alpha$ proportion of the codeword if there exist $n+1$ functions $f_i:F^{l}\to F^{\alpha_i l}, i=1,2,\dots,n$ with $\sum_{i=1}^n \alpha_i\le n\alpha$ and 
$g: F^{(\sum_{i=1}^n\alpha_i) l} \to F^{nl}$ such that
\begin{equation}\label{eq:gf}
g ( f_1(C_1+E_1), f_2(C_2+E_2),\dots,f_n(C_n+E_n) )=(C_1,C_2,\dots,C_n)
\end{equation}
for any codeword $(C_1,\dots,C_n)\in\cC$ and any error vector $(E_1, E_2, \dots, E_n)$ of Hamming weight $|\{i:E_i\neq 0\}|\le t.$

(ii) For $\alpha \ge k/n,$ define the $\alpha$-decoding radius $r_{\alpha}(\cC)$ as the maximum number of errors that the code
$\cC$ can correct by downloading an $\alpha$-proportion of the codeword.
%

(iii) For $\alpha \ge k/n,$ we further define the $\alpha$-decoding radius of $(n,k)$ codes as $$
r_{\alpha}(n,k)=\max_{\cC\in\cM_{n,k}}r_{\alpha}(\cC),
$$
where $\cM_{n,k}$ is the set of all $(n,k)$ codes.
\end{definition}
\begin{remark} Since the information contents of the codeword $C$ is $kl$ symbols of the field $F$, the inequality
$\alpha\ge k/n$ forms a trivial necessary condition for decoding even without errors. This condition will be assumed throughout
the paper.
\end{remark}

It is well known that for any $(n.k)$ code $\cC,$ we have $r_1(\cC)\le \lfloor(n-k)/2\rfloor,$ and the equality holds for MDS codes. Thus $R_1(n,k)=\lfloor(n-k)/2\rfloor.$ Moreover, we have an obvious lower bound for an MDS code $\cC$:
\begin{equation}\label{eq:LB}
r_{\alpha}(\cC) \ge \lfloor(\alpha n-k)/2\rfloor.
\end{equation}
 To see this, we can simply read any $\alpha n$ coordinates of the codeword. Since a punctured MDS code is still an MDS code with the same dimension, we obtain the lower bound \eqref{eq:LB}.

In this paper, we show that 
\begin{equation}\label{eq:RA}
r_{\alpha}(n,k)=\lfloor(n-k/\alpha)/2\rfloor
\end{equation}
 for any $n,k$ and $\alpha,$ and we give two families of explicit constructions of MDS codes together with the decoding schemes which achieve the optimal $\alpha$-decoding radius in \eqref{eq:RA}. The optimal $\alpha$-decoding radius in \eqref{eq:RA} improves upon the lower bound \eqref{eq:LB} obtained from the naive decoding strategy by a factor of $1/\alpha.$

The underlying idea of the two optimal code constructions and decoding schemes is to download from each of the codeword
coordinates a number of field symbols that forms an $\alpha$ proportion of the coordinate's size, and to 
ensure that the downloaded symbols constitute a codeword in an $(n, k/\alpha, \alpha l)$ MDS code, which can be used to recover the original data. One of our constructions is a short Reed-Solomon code with evaluation points in a subfield, and the other
is based on Folded Reed-Solomon (FRS) codes of {\sc Guruswami} and {\sc Rudra} \cite{Guruswami08}.
While FRS codes solve the problem somewhat trivially, our construction of the short RS code has the advantage of smaller node (codeword coordinate) size.

Furthermore, we show that random codes with high probability asymptotically achieve the bound \eqref{eq:RA} on the $\alpha$-decoding radius. The ensemble of random codes that we consider is based on randomly chosen ``contracting'' linear maps of the coordinates of an MDS code. 
Finally, we take up the question of constructing MDS codes with optimal repair bandwidth (also called MSR codes) which at the same
time have the optimal $\alpha$-decoding radius. A construction of codes with both these properties is obtained by using an
idea in a recent paper \cite{Ye16} by the authors.

The paper is organized as follows. In Section~\ref{sectub} we prove an upper bound on the $\alpha$-decoding radius, which we show to
be attainable in several ways. Specifically, in Section~\ref{sect:rcb} we show that random linear mappings are asymptotically optimal for fractional decoding. Subsequently, in Sections~\ref{constr1} and \ref{constr2} we present the two families of code constructions achieving the upper bound for finite code length.
Then in Section~\ref{constr2} we introduce the notion of $\alpha$-list decoding capacity, and show that both code constructions achieve the $\alpha$-list decoding capacity.
Finally, in Section~\ref{sect:msr}, we present the MSR code construction with optimal $\alpha$-decoding radius.

\section{upper bound on the $\alpha$-decoding radius}\label{sectub}

\begin{theorem}\label{thm:r_alpha} Let $n\ge \alpha n \ge k.$ Then
\begin{equation}\label{eq:ub}
r_{\alpha}(n,k)\le \lfloor (n-k/\alpha)/2\rfloor.
\end{equation}
\end{theorem}

\begin{proof}
Let $\cC$ be an $(n,k,l)$ code. Consider the ``projected'' code $C^\alpha$ obtained by applying the functions $f_i,i=1,\dots,n$ to the coordinates of the codewords of $\cC$:
   $$
   \cC^\alpha=\{(f_1(C_1),...,f_n(C_n)):(C_1,...,C_n)\in \cC\}
   $$
We will argue that the distance of the code $\cC^\alpha$ is at most $n-\lceil \frac{k}{\alpha}\rceil+1,$ implying \eqref{eq:ub}.
Suppose otherwise, then the code $\cC^\alpha$ corrects any $n-\lceil \frac{k}{\alpha}\rceil+1$ erasures, i.e., it is possible
to recover the codeword from any given subset of $s:=\lceil \frac{k}{\alpha}\rceil-1$ of its coordinates.

Assume w.l.o.g. that $\alpha_1\le\alpha_2\dots\le\alpha_n.$ By the assumption, it is possible to recover the codeword from the first $s$ coordinates, i.e., the projection mapping on the first $s$ coordinates is
injective, or, rephrasing again, $\sum_{i=1}^s\alpha_i\ge k.$ This implies that $\alpha_{s+1}\ge\alpha_s\ge k/s.$ With this we obtain
   $$
   \sum_{i=1}^n\alpha_i=\sum_{i=1}^s\alpha_i+\sum_{i=s+1}^n\alpha_i\ge k + (n-s)\frac ks=\frac{nk}s> \alpha n
   $$
since $s<\frac k\alpha.$   At the same time, by Def.~\ref{def:radius}, the sum $\sum_{i=1}^n\alpha_i\le \alpha n$, which contradicts the assumption.
The proof is complete.

%
\end{proof}

\section{Random coding bounds} \label{sect:rcb}
Here we examine another view of the codes defined above with the aim of estimating the parameters of codes $\cC^\alpha$
obtained from MDS codes under a random contracting mapping. To put the arguments in context, recall the construction of concatenated codes which combine two codes, say an $[n,k]$ MDS code $\cC_1$ over the finite field $F=\mathbb{F}_{q^l}$ and an $[m,l]$ code $\cC_2$ over the field $\mathbb{F}_q,$ into a code of length $nm$ over $\mathbb{F}_q.$ To transform a codeword $C=(C_1,\dots, C_n)$ of $\cC_1$ to the codeword of the concatenated code, each symbol $C_i$ is replaced with a codeword of the code $\cC_2$ using some injective
map from $\mathbb{F}_{q^l}$ to $\cC_2.$ Thereby, the number of coordinates in the $q$-ary representation of $C_i$ is increased from $l$ to $m$.
In our current situation, we are interested in the code obtained by mapping the coordinate $C_i$ to an element in the field 
$\mathbb{F}_{q^{\alpha l}},$ where $\alpha<1$ (Definition \ref{def:radius} considers a slightly more general case wherein
$\alpha$ depends on $i$, while the construction of the next section assumes equal $\alpha_i$'s). Thus, codes for fractional decoding
may be viewed as ``inverse concatenation codes'' which shrink the dimension of each coordinate of the original codes instead of expanding it.

This point of view suggests an approach to random coding bounds similar to the earlier results on concatenated codes e.g., \cite{tho83}.
Namely, we start with an $[n,k]$ MDS code $\cC_1$ over the field $\mathbb{F}_{q^l}$ and map each coordinate to an element in $\mathbb{F}_{q^{\alpha l}}$ using a uniformly random linear mapping. Specifically, suppose that $A=(A_1,\dots,A_n)$ is an $n$-tuple of linear maps $\mathbb{F}_{q^l}\to \mathbb{F}_{q^{\alpha l}}$  and let 
  $$\cC^\alpha:=A(\cC_1){ =\{(A_1(C_1),\dots,A_n(C_n)):(C_1,\dots,C_n)\in\cC_1\} }$$ 
  be the resulting linear code. In this section
we compute the typical parameters of the code $\cC^\alpha,$ which will be shown to meet the bound
\eqref{eq:RA} with high probability. We consider two different asymptotic regimes, of fixed $n$ and $l\to\infty,$ and of
$n=q^l\to \infty,$ with the above conclusion applying to both of them.

We will call the mapping $A$ {\em optimal} for the fractional decoding of $\cC_1$ if for every subset $\cI\subset [n]$ of size 
$L=k/\alpha+1,$ the restriction of $A$ to $\cI$ defined as 
   \begin{equation}\label{eq:opt}
   \begin{aligned}
   A_{\cI}: \quad\quad\quad\quad \; {\cC_1} &\to \left(\mathbb{F}_{q^{\alpha l}}\right)^{L}\\
(C_1,\dots,C_n) & \mapsto  (A_i(C_i),i\in\cI)
  \end{aligned}
  \end{equation}
is injective. Recalling Definition~\ref{def:radius} and the bound \eqref{eq:RA}, if $A$ is optimal, then the code
$\cC^\alpha$ corrects $n-L$ erasures, and so its distance equals $n-k/\alpha.$ 
%
%
Suppose that $A=(A_1,\dots,A_n)$ is realized by random $l\times \alpha l$ matrices $A_i$
whose elements are chosen from $\mathbb{F}_{q}$  independently and with uniform distribution.


 Before proceeding, recall the following classic fact about the weight distribution of an $[n,k]$ MDS code $\cC_1$ over $\mathbb{F}_{q^l}:$
    $$
    |\{C\in \cC_1: \wt(C)=i\}|\le \binom{n}{i} q^{l(i-n+k)},\;\; i\ge n-k+1.
    $$
Indeed, the restriction of $\cC_1$ to any $k$ coordinates is injective. Once we fix $n-i$ coordinates to 0 in any of the possible 
$\binom ni$ ways, there are $(q^l-1)^{k-(n-i)}$ possible choices of nonzero coordinates before the codeword is identified uniquely. This 
gives the claimed upper bound.

\begin{proposition}\label{prop:31}
 Let $\cC_1$ be an $[n,k]$ MDS code over the field $\mathbb{F}_{q^l}$. Let $\alpha>k/n$ and let
$A: \cC_1\to \cC^{\alpha}$ be the random linear mapping defined above.
Suppose that $n,k$ are fixed and $l\to\infty$, then $A$ is an optimal mapping for the fractional decoding of $\cC_1$ with probability $1-o(1)$.
\end{proposition}
\begin{proof}

Let $C=(C_1,\dots,C_n),C\neq 0$ be a codeword of $\cC_1$ and suppose its Hamming weight is $\wt(C)=w$. 
Since $A=(A_1,\dots,A_n)$ is linear, $A_i(C_i)=0$ if $C_i=0$ and $\Pr(A_i(C_i)=0)=q^{-\alpha l}$  if $C_i\neq 0$.
Therefore
$$
  \Pr(A(C)=0)=q^{-\alpha wl}.
$$

Observe that for any subset $\cI\subset [n]$ of size $L>k$, the code $\cC_1$ restricted to the coordinates in $\cI$ 
is an $[L,k]$ MDS code. Now let us fix a subset $\cI\subseteq[n]$ of size ${L}>k/\alpha$ and show that the mapping $A:\cC_1\rightarrow \cC^\alpha$
with high probability has a trivial kernel. We have
   \begin{align}
\Pr(\ker(A_\cI)\ne 0\})
& \le  \sum_{C\in\cC_{\cI}, C\neq 0} \Pr(A_\cI(C)=0) \nonumber\\
&=  \sum_{w={L}-k+1}^{{L}} \sum_{\wt(C)=w} \Pr(A_\cI(C)=0)  \nonumber\\
&\le  \sum_{w={L}-k+1}^{{L}} \binom{{L}}{w} q^{l(w-{L}+k)} q^{-\alpha wl}  \nonumber\\
&=  \sum_{w={L}-k+1}^{{L}} \binom{{L}}{w} q^{l(w-\alpha w-{L}+k)}.\label{eq:jp}
\end{align}
The exponent in the last expression, given by $w-\alpha w-{L}+k,$ is an increasing function of $w$, so 
$w-\alpha w-{L}+k \le k-\alpha{L}<0$ for all $w\le {L}$.
Therefore $q^{l(w-\alpha w-{L}+k)} \to 0$ for all $w\le {L}$ when $l\to\infty,$ and thus $\Pr(\ker(A_\cI)\ne 0\})\to 0$
 for every subset $\cI\subseteq[n]$ of size ${L}>k/\alpha$. Since there are only finitely many such subsets, we conclude that with probability approaching one, the mapping $A_\cI$ is injective for every choice of $\cI$. This completes the proof of the proposition.
\end{proof}

Now let us analyze the case when the code length $n=q^l\to\infty.$ In this case it is more convenient to consider asymptotic
optimality of the mapping $A.$ Given 
an $[n,k=Rn]$ MDS code $\cC_1$ over the field $\mathbb{F}_{q^l}$ and a linear mapping 
$A: \cC_1\mapsto\cC^{\alpha}$, we call $A$ {\em asymptotically optimal} for the fractional decoding of $\cC_1$ if the following two conditions are satisfied:
\begin{enumerate}
\item $A$ is injective;
\item the distance of the code $\cC^\alpha$ satisfies $d(\cC^{\alpha})\ge n(1-R/\alpha-o(1))$.
\end{enumerate}
In other words, the mapping $A$ is asymptotically optimal if the cardinality of the code $C^\alpha$ is unchanged from that of $\cC_1$, and its relative 
distance asymptotically satisfies the bound \eqref{eq:RA}.

\begin{proposition}\label{prop:32}
 Let $\cC_1$ be an $[n,k]$ MDS code over $\mathbb{F}_{q^l},$ where $n=q^l$ and $k=Rn$. 
Let $A=(A_1,\dots,A_n)$ be the random linear mapping $\cC_1 \to\cC^{\alpha}$ defined above, where $\alpha>R$.
Suppose that $R$ is fixed and $n\to\infty$, then $A$ is an asymptotically optimal mapping for the fractional decoding of $\cC_1$ with probability $1-o(1)$.
\end{proposition}
\begin{proof}
Let us prove the injectivity condition. Proceeding as in \eqref{eq:jp}, we have
\begin{align*}
\Pr(\ker(A)\ne 0\})
& \le\sum_{C\in\cC_1, C\neq 0} \Pr(A(C)=0) 
=  \sum_{w=n-k+1}^{n} \sum_{\wt(C)=w} \Pr(A(C)=0) \\
&\le  \sum_{w=n-k+1}^{n} \binom{n}{w} q^{l(w-n+k)} q^{-\alpha wl} 
= q^{-nl(1-R)} \sum_{w=n-k+1}^{n} \binom{n}{w} q^{wl(1-\alpha)} \\
&  \le  q^{-nl(1-R)} \sum_{w=0}^{n} \binom{n}{w} q^{wl(1-\alpha)}
 =q^{-nl(1-R)} (1+q^{l(1-\alpha)} )^n \\
& = (q^{-l(1-R)} + q^{-l(\alpha-R)} )^n
\to 0.
\end{align*}
This shows that the mapping $A$ is injective with probability $1-o(1)$.

Next we prove that with probability $1-o(1)$ the distance $d(\cC^{\alpha})$ satisfies
 \begin{equation}\label{eq:d}
  d(\cC^\alpha)\ge n-\frac{k}{\alpha} -\frac{2n}{\alpha\log_4 n}=n\Big(1-\frac{R}{\alpha}-o(1)\Big).
 \end{equation}
Starting with a nonzero codeword $C\in\cC_1$ of weight $\wt(C)=w$, let us estimate the probability
that it maps on a codeword of $\cC^\alpha$ of weight no larger than $i$ for some $i\le w:$
$$
\Pr(\wt(A(C))\le i) \le \binom{w}{i} q^{-\alpha l(w-i)}.
$$
By the union bound, 
\begin{align*}
\Pr(d(C^\alpha)\le i)
&\le \Pr(\{\exists C\in\cC_1: 1\le \wt(A(C))\le i\}) \\
&\le \sum_{C\in\cC_1, C\neq 0} \Pr(\wt(A(C)) \le i) \\
&=  \sum_{w=n-k+1}^{n} \sum_{\wt(C)=w} \Pr(\wt(A(C)) \le i)  \\
&\le \sum_{w=n-k+1}^{n} \binom{n}{w} q^{l(w-n+k)} \binom{w}{i} q^{-\alpha l(w-i)} \\
&=  \sum_{w=n-k+1}^{n} \binom{n}{w}  \binom{w}{i} n^{w-\alpha w-n+k+ \alpha i}\\
&\overset{(a)}{\le} \sum_{w=n-k+1}^{n} 4^n n^{-\alpha n+k+\alpha i}\\
&\le k 4^{n+ (k-\alpha n+\alpha i)\log_4 n},
\end{align*}
where inequality $(a)$ follows from the facts that $\binom{n}{w}\le 2^n, \binom{w}{i}\le 2^n,$ and $w-\alpha w-n+k+ \alpha i<-\alpha n+k+\alpha i$ for all $w\le n$.
Thus if $i=n-\frac{k}{\alpha} -\frac{2n}{\alpha\log_4 n}$, then
$$
\Pr(d(C^\alpha)\le i) 
\le 4^{-n} k  \to 0
$$
when $n\to\infty$. 
This implies \eqref{eq:d} and concludes the proof.
\end{proof}

Concluding this section, we note the difference between the results for classic binary concatenated codes \cite{tho83} and the results 
above. In the former case, symbols of the MDS code are mapped on random binary codewords, and the resulting code with high probability
approaches the Gilbert-Varshamov bound, matching the best known parameters for the binary case (under some 
additional assumption on the component codes, derived in \cite{tho83}.) In our case, the alphabet of the resulting
code $\cC^\alpha$ is allowed to grow, and the rate and distance of $\cC^\alpha$ are as good as those obtained from MDS codes in
a deterministic way {in the next two sections}.


\section{Construction of RS codes that are optimal for fractional decoding}\label{constr1}
All the constructions in this paper derive from the RS code family, defined as follows.
\begin{definition}
A \emph{Reed-Solomon code} $\text{\rm RS}_F(n,k,\Omega)\subseteq F^n$ of dimension $k$ over $F$ 
with evaluation points $\Omega=\{\omega_1,\omega_2,\dots,\omega_n\}\subseteq F$  is the set of vectors
$$
\{(h(\omega_1),\dots,h(\omega_n))\in F^n:h\in F[x], \deg h\le k-1\}.
$$ 
\end{definition}

In this section we construct a family of RS codes with carefully chosen evaluation points achieving optimal $\alpha$-decoding radius
(we assume throughout that $\alpha$ is rational, noting that this constraint does not incur any loss of generality in terms of the code parameters).
We will use the  field trace function, which is defined as
\begin{definition}
Let $F=\mathbb{F}_{q^s}$ be a finite field extension of $B=\mathbb{F}_q$ of degree $s.$ The field trace is defined as
$$
\tr_{F/B}(\beta)=\beta+\beta^q+\beta^{q^2}+\dots+\beta^{q^{s-1}}.
$$
Let $\zeta_0, \zeta_1, \dots, \zeta_{s-1}$ be a basis of $F$ over $B,$ and let $\nu_0, \nu_1,\dots, \nu_{s-1}$ be the dual basis, then
$$
\beta=\sum_{i=0}^{s-1} \tr_{F/B}(\zeta_i\beta)\nu_i.
$$
In other words, any element $\beta$ in $F$ can be calculated from its $s$ projections $\{\tr_{F/B}(\zeta_i\beta)\}_{i=0}^{s-1}$ on $B.$
\end{definition}

\begin{proposition}\label{prop:codes-subfield}
Let $\alpha=m/s < 1,$ where $m$ and $s$ are positive integers. 
Given $n$ and $k$ satisfying that $n\ge sk/m$ and $m|k,$ let $F=\mathbb{F}_{q^s}, q\ge n$ be a finite field extension of $B=\mathbb{F}_q$ of degree $s.$ An $(n,k)$ code $\text{\rm RS}_F(n,k,\Omega)\subseteq F^n$ with all the evaluation points $\Omega=\{\omega_1,\omega_2,\dots,\omega_n\}\subseteq B$ has the optimal $\alpha$-decoding radius.
\end{proposition}
The proof is given in the remainder of this section.
Each codeword coordinate is a vector of dimension $s$ over $B.$ Thus $\text{\rm RS}_F(n,k,\Omega)$ can be viewed as an $(n,k,s)$ MDS array code over the base field $B.$
 Our strategy is to download $m$ symbols in $B$ from each of the codeword coordinate, which is exactly $m/s$ proportion of the codeword.

Before explaining which $m$ symbols in $B$ we download from each of the coordinates, let us introduce some notation.
We write the encoding polynomial as
\begin{equation}\label{eq:defh}
h(x)=a_{k-1}x^{k-1}+a_{k-2}x^{k-2}+\dots+a_0.
\end{equation}
Then the $i$-th coordinate of the codeword is 
\begin{equation}\label{eq:ci}
c_i=h(\omega_i)=a_{k-1}\omega_i^{k-1}+a_{k-2}\omega_i^{k-2}+\dots+a_0.
\end{equation}
Let $\zeta_0, \zeta_1, \dots, \zeta_{s-1}$ be a basis of $F$ over $B.$
For $j=0,1,\dots,s-1,$ we further define 
$$
h_j(x)=\tr_{F/B}(\zeta_j a_{k-1}) x^{k-1} + \tr_{F/B}(\zeta_j a_{k-2}) x^{k-2} +\dots
+ \tr_{F/B}(\zeta_j a_0).
$$
Since the coefficients of $\{h_j(x)\}_{j=0}^{s-1}$ contain all the projections of the coefficients of $h(x)$ onto $B,$ the coefficients of $h(x)$ can be calculated from the coefficients of $\{h_j(x)\}_{j=0}^{s-1}.$
In other words, to recover the codeword, it suffices to know $\{h_j(x)\}_{j=0}^{s-1}.$

Let $A_0,A_1,\dots,A_{m-1} \subseteq B$ be $m$ pairwise disjoint subsets of the field $B,$ each of size $k/m.$
For $j=0,1,\dots,m-1,$ define the annihilator polynomials of the set $A_j$ to be
$$
p_j(x) = \prod_{\omega \in A_j}(x-\omega).
$$

The $m$ symbols we download from the $i$-th coordinate are as follows:
\begin{equation}\label{eq:dl}
d_i^{(j)}= \tr_{F/B}(\zeta_{s-m+j} c_i) (p_j(\omega_i))^{s-m} +
\sum_{u=0}^{s-m-1} \tr_{F/B}(\zeta_{u} c_i) (p_j(\omega_i))^u, \quad 
j=0,1,\dots,m-1.
\end{equation}
Clearly, $d_i^{(j)}\in B$ for all $j=0,1,\dots,m-1.$
Plugging \eqref{eq:ci} into \eqref{eq:dl}, we can see that
$$
d_i^{(j)}= g_j(\omega_i), \quad 
j=0,1,\dots,m-1.
$$
where
\begin{equation}
\label{eq:stam}
g_j(x)=h_{s-m+j}(x) (p_j(x))^{s-m} + \sum_{u=0}^{s-m-1} h_u(x) (p_j(x))^u, \quad 
j=0,1,\dots,m-1.
\end{equation}

Since $\deg(p_j)=k/m,$ we have $\deg(g_j)<sk/m.$ Thus $(d_1^{(j)},d_2^{(j)},\dots,d_n^{(j)})\in\text{\rm RS}_B(n,sk/m,\Omega)$ for every $j=0,1,\dots,m-1.$ As a result, we can recover all the coefficients of polynomials $\{g_j(x)\}_{j=0}^{m-1}$ as long as there are no more than $\lfloor(n-sk/m)/2\rfloor$ errors in the original codeword $(c_1,c_2,\dots,c_n).$ 
Now we only need to show that given polynomials $\{g_j(x)\}_{j=0}^{m-1},$ we can recover the polynomials $\{h_j(x)\}_{j=0}^{s-1}.$ To see this, we notice that for $j=0,1,\dots,m-1,$
$$
g_j(\omega)=h_0(\omega) \text{~for all~}\omega \in A_j.
$$
Consequently, we know the evaluations of $h_0(x)$ at all the points in $\cup_{j=0}^{m-1}A_j.$ There are $k$ distinct points in the set $\cup_{j=0}^{m-1}A_j$ and the degree of $h_0(x)$ is less than $k,$ so we can recover $h_0(x).$ From $h_0(x)$ and $\{g_j(x)\}_{j=0}^{m-1},$ we can calculate the polynomials
$$
g'_j(x)=\frac{g_j(x)-h_0(x)}{p_j(x)}= h_{s-m+j}(x) (p_j(x))^{s-m-1} + \sum_{u=1}^{s-m-1} h_u(x) (p_j(x))^{u-1}, \quad 
j=0,1,\dots,m-1.
$$
Since 
$$
g'_j(\omega)=h_1(\omega) \text{~for all~}\omega \in A_j,
$$
we know the evaluations of $h_1(x)$ at all the points in $\cup_{j=0}^{m-1}A_j.$ So we can also recover $h_1(x).$ From $h_0(x), h_1(x)$ and $\{g_j(x)\}_{j=0}^{m-1},$ we can calculate the polynomials
$$
g''_j(x)=\frac{g'_j(x)-h_1(x)}{p_j(x)}= h_{s-m+j}(x) (p_j(x))^{s-m-2} + \sum_{u=2}^{s-m-1} h_u(x) (p_j(x))^{u-2}, \quad 
j=0,1,\dots,m-1.
$$
Since 
$$
g''_j(\omega)=h_2(\omega) \text{~for all~}\omega \in A_j,
$$
we know the evaluations of $h_2(x)$ at all the points in $\cup_{j=0}^{m-1}A_j.$ So we can also recover $h_2(x).$ It is clear that we can repeat this procedure until we recover $\{h_j(x)\}_{j=0}^{s-m-1}.$
Then the polynomials $\{h_{s-m+j}(x)\}_{j=0}^{m-1}$ can be easily recovered by
$$
h_{s-m+j}(x)  =\frac{ g_j(x)- \sum_{u=0}^{s-m-1} h_u(x) (p_j(x))^u} {(p_j(x))^{s-m}}, \quad 
j=0,1,\dots,m-1.
$$
This shows that we can recover the polynomials $\{h_j(x)\}_{j=0}^{s-1}$ from the polynomials $\{g_j(x)\}_{j=0}^{m-1},$ and consequently recover the original codeword.

\section{Folded Reed-Solomon codes}\label{constr2}
Folded RS (FRS) codes were introduced by {\sc Guruswami} and {\sc Rudra} \cite{Guruswami08}
for the problem of optimal list decoding. In this section we show that FRS codes are optimal for the fractional decoding in
a rather straightforward way.

Let us recall the definition of FRS codes.

\begin{definition}
Let $F$ be a finite field with cardinality $|F| > nl.$ Let $\gamma$ be a primitive element of $F.$
A Folded Reed-Solomon code $\text{\rm FRS}(n,k,l)\subseteq (F^{l})^n$ is an MDS array code with each codeword coordinate being a vector in $F^l$ defined as follows:
\begin{align*}
\{(C_1, C_2, \dots, C_n) : C_i= &(h(\gamma^{(i-1)l}), h(\gamma^{(i-1)l+1}), \dots, h(\gamma^{(i-1)l+l-1})\in F^l \text{ for } 1\le i\le n,\\
 &h\in F[x], \deg h\le kl-1 \}.
\end{align*}
\end{definition}

We limit ourselves to those values of sub-packetization $l$ for which $\alpha l$ is an integer.
\begin{proposition} The $\alpha$-decoding radius of FRS codes satisfies
  $$
  r_{\alpha}(\text{\rm FRS}(n,k,l))= \lfloor(n-k/\alpha)/2\rfloor.
  $$
\end{proposition}  
\begin{proof} We will construct $n+1$ functions $f_i:F^{l}\to F^{\alpha_i l}, i=1,2,\dots,n$ and $g: F^{(\sum_{i=1}^n\alpha_i) l} \to F^{nl}$ that \eqref{eq:gf}. The functions $f_i:F^{l}\to F^{\alpha l}$ will simply project a symbol on its first $\alpha l$ coordinates, i.e., $f_i=f,$
where for $(d_1,d_2,\dots,d_l)\in F^{l},$
\begin{equation}\label{eq:deff}
f((d_1,d_2,\dots,d_l))=(d_1, d_2, \dots, d_{\alpha l}).
\end{equation}
Thus, the code $\cC^\alpha$ is a projection of the code $\cC$, 
\begin{equation}\label{eq:Calpha1}
\cC^{\alpha}=\{(C_1^{\alpha},C_2^{\alpha},\dots,C_n^{\alpha}) = (f(C_1), f(C_2),\dots,f(C_n)) : (C_1,C_2,\dots,C_n)\in\text{\rm FRS}(n,k,l)\}
\end{equation}
Equivalently, we can write $C^{\alpha}$ as
\begin{equation}\label{eq:Calpha2}
\begin{aligned}
C^{\alpha}=\{(C_1^{\alpha}, C_2^{\alpha}, \dots, C_n^{\alpha}) : C_i^{\alpha}= &(h(\gamma^{(i-1)l}), h(\gamma^{(i-1)l+1}), \dots, h(\gamma^{(i-1)l+\alpha l-1})\in F^l \text{ for } 1\le i\le n,\\
 &h\in F[x], \deg h\le kl-1 \}.
\end{aligned}
\end{equation}
Since any $k/\alpha$ coordinates of $\cC^{\alpha}$ contain $(k/\alpha)(\alpha l)$ evaluations of the encoding polynomial $h$ with degree less than $kl,$ we can recover $h$ and thus the whole codeword from any $k/\alpha$ coordinates of $\cC^{\alpha}.$ We thus conclude that $\cC^{\alpha}$ is an $(n,k/\alpha,\alpha l)$ MDS array code, so it can correct up to $\lfloor(n-k/\alpha)/2\rfloor$ errors. 

If $E_i$ is the error in the $i$th coordinate of the codeword, we can write $f(C_i+E_i)=f(C_i)+f(E_i)$ for $i=1,2,\dots,n.$
Suppose that $(C_1,C_2,\dots,C_n)\in\text{\rm FRS}(n,k,l)$ and $|\{i:E_i\neq 0\}|\le \lfloor(n-k/\alpha)/2\rfloor,$ then $(f(C_1), f(C_2),\dots, f(C_n))\in\cC^{\alpha}$ and
$|\{i:f(E_i)\neq 0\}|\le \lfloor(n-k/\alpha)/2\rfloor.$
As a result, we can recover the codeword $(f(C_1), f(C_2),\dots, f(C_n))\in\cC^{\alpha}$ and thus
recover the encoding polynomial $h$ and finally the codeword $(C_1,C_2,\dots,C_n)\in\text{\rm FRS}(n,k,l)$ from $(f(C_1+E_1), f(C_2+E_2),\dots, f(C_n+E_n)).$ 
By our definition in \eqref{eq:gf}, this shows that $r_{\alpha}(\text{\rm FRS}(n,k,l))\ge \lfloor(n-k/\alpha)/2\rfloor,$
and proof is concluded with a reference to the upper bound \eqref{eq:ub}.
%
%
\end{proof}

\begin{remark}
Given multiple values $\alpha_1, \alpha_2, \dots, \alpha_m,$ if we choose $l$ in such a way that $\alpha_1 l, \alpha_2 l,\dots, \alpha_m l$ are all integers, then $\text{\rm FRS}(n,k,l)$
achieves the optimal $\alpha_i$-decoding radius for $1\le i\le m$ simultaneously.
\end{remark}

We can use the decoding method described above to give more general code constructions achieving the optimal $\alpha$-decoding radius. 
Indeed, we can take any $(nl, kl)$ scalar MDS code over a finite field $F$ and group
together blocks of $l$ coordinates of it into a vector in $F^l.$ 
It is clear that in this way we obtain an $(n,k,l)$ MDS array code $\tilde\cC$. 
Moreover, by reading $\alpha l$ symbols of $F$ from each of the coordinates of $\tilde\cC$ we 
obtain an $(n,k/\alpha,\alpha l)$ MDS array code $\cC^{\alpha}$ which can correct up to 
$\lfloor(n-k/\alpha)/2\rfloor$ errors, and thus $\cC$ forms an optimal code for fractional decoding.

\begin{remark}\label{remark:ell}
We note that the RS codes of Section~\ref{constr1} are somewhat preferable to FRS codes because in the context of the problem considered. Indeed, although both families have the largest possible $\alpha$-decoding radius, the FRS codes require larger node (codeword coordinate) size. Namely, 
the base field in the RS construction (the field $B$ in Section~\ref{constr1}) only needs to have size $n,$ 
while the base field in this section must be of size $nl.$ Consequently, the size of the codeword coordinate for the RS construction is $l\log n$ bits, while for FRS code it is $l\log (nl)$ bits. (Note that in the RS construction $l$ is the degree of the field extension, which is denoted by $s$ in Section~\ref{constr1}.)
\end{remark}


\section{$\alpha$-List Decoding Capacity}
In this section we extend our study to the list decoding problem. 
Under unique decoding, the decoder outputs the correct codeword as long as the received vector is within a certain distance $r_u$ 
from it. 
Under list decoding, the decoder finds a list of all codewords that are within a certain distance $r_l$ from the received vector.
Denote the size of this list by $L.$  We say that a code corrects $r_l$ errors under list-of-$L$ decoding if  sphere of radius $r_l$ centered at any received vector contains at most $L$ codewords. 

Complexity considerations suggest that $L$ is a slowly growing function of the code length $n$
(or even a constant). In this paper, following a long line of work in algebraic list decoding, we assume that $L$ is a polynomial 
function of $n$. The main result of \cite{Guruswami08} amounts to stating that $(n,k,l)$ FRS codes of rate $R:=k/n$ correct the asymptotically maximum number of errors $r_l=n(1-R+o(1))$ under lists of polynomial size.
It turns out that FRS codes are also optimal under fractional list decoding.


Let us define formally the fractional decoding problem.
\begin{definition}[$(\alpha,L)$ list decoding radius]\label{def:listradius} 
Consider an $(n,k,l)$  array code $\cC=\{(C_1,\dots,C_n)\}$ over $F$, 
where $C_i\in F^l, i=1,\dots, n$. 

(i) We say that $\cC$ corrects up to $t$ errors under list-of-$L$ decoding by downloading an $\alpha$ proportion of the codeword if there exist $n+1$ functions $f_i:F^{l}\to F^{\alpha_i l}, i=1,2,\dots,n,$ $\sum_{i=1}^n \alpha_i \le n\alpha$ and 
$g: F^{(\sum_{i=1}^n\alpha_i) l} \to (F^{nl})^L$ such that for any codeword $C=(C_1,\dots,C_n)\in\cC$ and any error vector $E=(E_1, E_2, \dots, E_n)$ of Hamming weight $\le t,$ we have
\begin{equation}\label{eq:listgf}
\begin{aligned}
g ( f_1(C_1+E_1),\;  &f_2(C_2+E_2),\dots,f_n(C_n+E_n) )=\{C^{(i)}, i=1,\dots,L\},\\
&\text{and } C\in \{C^{(i)}, i=1,\dots,L\}.
\end{aligned}
\end{equation}

(ii) For $\alpha \ge k/n,$ define the $(\alpha, L)$-list decoding radius $r_{\alpha,L}(\cC)$ as the maximum number of errors that the code $\cC$
can correct by downloading an $\alpha$ proportion of the codeword.

(iii) For $\alpha \ge R,$ we further define the (normalized) $\alpha$-list decoding capacity of codes of rate at least $R$ as
$$
\rho_{\alpha}(R)=\sup \Big\{\frac{r_{\alpha,L}(\cC)}{n}: \text{rate}(\cC)\ge R \text{ and } L \text{ is polynomial in }n\Big\},
$$
where $n(\cC)$ is the code length of $\cC.$ More formally, 
  $$
\rho_{\alpha}(R)=\sup_{m\in {\mathbb N}}\limsup_{n\to\infty} \frac{r_{\alpha,n^m}(n,Rn)}{n}
$$
where $r_{\alpha,n^m}(n,Rn)$ is the maximum of $r_{\alpha,L}(\cC)$ over all codes of length $n$ and rate $R$.
\end{definition}

Repeating the proof of Theorem~\ref{thm:r_alpha}, we can easily show that $\rho_{\alpha}(R)\le 1-R/\alpha.$ 
At the same time, we can show that the two families of RS-type codes shown above to be optimal for $\alpha$-decoding
are also optimal for the fractional list decoding problem in the sense of achieving the $\alpha$-list decoding
capacity.

\subsection{$\alpha$-List decoding the codes in Sec.~\ref{constr1}}
We recall that the codes in Prop.~\ref{prop:codes-subfield} are simply RS codes with 
evaluation points in a subfield. Such codes have appeared in several previous works on array codes; in particular, 
in \cite{Guruswami-Xing}, {\sc Guruswami} and {\sc Xing} presented
a list decoding algorithm for them. This algorithm can be easily modified for the problem of $\alpha$-list decoding the
 codes of Prop.~\ref{prop:codes-subfield}.
   \begin{theorem}[\cite{Guruswami-Xing}, Corollary 4.5]
\label{stam3} Let $B=\mathbb{F}_q, F=\mathbb{F}_{q^s}$ and let $\cC$ be the code $\text{RS}_F(n,k,\Omega)$, where $\Omega=B.$
For every $R=\frac{k}{n}\in (0,1)$, and $\epsilon, \gamma >0$, there exists a sufficiently large positive integer $s$ such that the code 
    can be list decoded from a fraction of $1-R-\epsilon$ of errors in $|\cC|^\gamma$ time, outputting a list of size at most 
    $|\cC|^\gamma.$
\end{theorem}
This result can be modified for the $\alpha$-list decoding problem, where as before $\alpha=k/m.$ 
The idea is to lift the vector formed from the downloaded symbols of $B$ back to $F$ and to use the algorithm of
Theorem \ref{stam3} for the RS code over $F$.

Let $\zeta_0,...,\zeta_{s-1}$ be a basis of $F$ over $B$. 
Following the procedure in Sec.~\ref{constr1}, we download symbols $d_i^{(j)}\in B, j=1,\dots,m$ 
from each of coordinates $i=1,\dots,n$ of the received vector as described in \eqref{eq:dl}.
Form the vector $(y_1,...,y_n)\in F^n$ where for $i=1,\dots,n$
     $$
     y_i:=\sum_{j=0}^{m-1}d_i^{(j)}\zeta_j
     $$
This vector can be viewed as a possibly corrupted version of the codeword
    \begin{equation}
      (G(\omega_1),...,G(\omega_q)), \text{ where } G(x):=\sum_{j=0}^{m-1}g_j(x)\zeta_j,
\label{stam4}
    \end{equation}
and where the polynomials $g_j(x)$ are defined in \eqref{eq:stam}. 
Since $\deg(g_j)<sk/m=k/\alpha,$ also $\deg(G)<k/\alpha$. Furthermore, the polynomial $G(x)\in F[x]$ 
is evaluated at the points of the subfield $B$, and thus 
 it can be viewed as a codeword of RS code of rate $k/(n\alpha)=R/\alpha
$. The list decoding algorithm outputs a list of codewords, which can be further pruned 
down by removing all polynomials $f\in F_{q^s}[x]$ that are not of the form \eqref{stam4}. 

This concludes the description, justifying the optimality claim for $\alpha$-list decoding of the codes considered here.

\subsection{$\alpha$-List decoding of FRS codes}
It is also possible to show that there exists a family of FRS codes
of growing length $n$ and sub-packetization $l$ that can be list-decoded from an $1-R/\alpha$ 
fraction of errors by downloading an $\alpha$ proportion of the codeword. To justify this claim,
we again need to construct $n+1$ functions $f_i:F^{l}\to F^{\alpha_i l}, i=1,2,\dots,n$ and 
$g: F^{(\sum_{i=1}^n\alpha_i) l} \to F^{nl}$ that satisfy \eqref{eq:listgf}. It turns out that
the projection functions suffice, and we take $f_1=f_2=\dots=f_n=f,$ where $f$ is defined in \eqref{eq:deff}. 
Downloading an $\alpha$ proportion from each of the codeword coordinates, we obtain the code $\cC^{\alpha}$ 
defined in \eqref{eq:Calpha2} whose rate is $R/\alpha.$ When the code length $n$ and sub-packetization $l$ of the FRS code become large enough, we can use the list decoding algorithm introduced in \cite{Guruswami08} to decode $\cC^{\alpha}$ up to a fraction arbitrarily close to $1-R/\alpha$ of errors.

Thus we conclude that
$$
\rho_{\alpha}(R)= 1-R/\alpha,
$$
and FRS codes achieve the $\alpha$-list decoding capacity.

\begin{remark}
The code $\cC^{\alpha}$ differs from an FRS code in the sense that the evaluation points in two consecutive coordinates are not consecutive powers of the primitive element.
However, the list decoding algorithm introduced in \cite{Guruswami08} only requires that within each codeword coordinate, the evaluation 
points are consecutive powers of the primitive element. The code $\cC^{\alpha}$ satisfies this constraint, so 
it is possible to rely on this algorithm in our arguments.
\end{remark}

Note that when the code length $n$ and the sub-packetization $l$ of FRS codes become large enough, they achieve the $\alpha$-list decoding capacity uniformly for all values of $\alpha.$ 
Note also that Remark~\ref{remark:ell} applies to the solutions of the 
list decoding problem that rely on RS codes of Section~\ref{constr1} and on FRS codes.

\section{Minimum Storage Regenerating codes with optimal $\alpha$-decoding radius}\label{sect:msr}
In this section we give an explicit construction of MDS codes with optimal bandwidth for repairing single erasure and optimal $\alpha$-decoding radius simultaneously. The construction is a simple extension of the MSR code construction in \cite{Ye16}.

We first recall the repair bandwidth and the cut-set bound.
Given an $(n,k,l)$ MDS array code $\cC$ over a finite field $F$, a failed node $C_i$ and a set of $d\ge k$ helper nodes $\{C_j,j\in\cR\}$, define $N(\cC,i,{\cR})$ as the smallest number of symbols of $F$ one needs to download  in order to recover the failed  
node $C_i$ from the helper nodes $\{C_j,j\in\cR\}$.
The repair bandwidth of the code is defined as follows.
\begin{definition}[Repair bandwidth]
Let $\cC$ be an $(n,k,l)$ {MDS} array code over a finite field $F$. Let $d\ge k$ be the number of helper nodes.
  The \emph{$d$-repair bandwidth} of the code $\cC$ is given by 
  \begin{equation}\label{eq:beta}
\beta(d):=\max_{i\in[n],|{\cR}|=d, i\notin\cR} N(\cC,i,{\cR}).
  \end{equation}
\end{definition}
According to the {\em cut-set bound} derived in \cite{Dimakis10},
$$
N(\cC,i,{\cR}) \ge \frac{dl}{d-k+1}
$$
for all $\cR\subseteq ([n]\setminus\{i\})$ with cardinality $d$.
If the $d$-repair bandwidth meets the cut-set bound with equality, i.e., 
$$
\beta(d) = \frac{dl}{d-k+1},
$$
 we say that the code $\cC$ has the {\em $d$-optimal repair property}, and $\cC$ is referred to as MSR code in the literature.

Let $\alpha=m/s < 1,$ where $m$ and $s$ are positive integers.
In this section we present an $(n,k,l=s(d-k+1)^n)$ MDS array code $\cC$ over a finite field $F$ with $d$-optimal repair property and optimal $\alpha$-decoding radius simultaneously, where the field size $|F|\ge s(d-k+1)n$.
We write a codeword of $\cC$ as $(C_1,C_2,\dots,C_n)$ and write each coordinate as 
$C_i=(c_{i,j,\underline{a}}:j\in[s],\underline{a}\in\{0,1,\dots,d-k\}^n)$, i.e., the coordinates of $C_i$ is indexed by a scalar $j\in[s]$ and a vector $\underline{a}=(a_1,a_2,\dots,a_n)\in\{0,1,\dots,d-k\}^n$, so each $C_i$ indeed has $l=s(d-k+1)^n$ coordinates.
Let $\{\lambda_{i,j,t}:i\in[n],j\in[s],t\in\{0,1,\dots,d-k\}\}$ be $s(d-k+1)n$ distinct elements of $F$.
The code $\cC$ is defined by the following set of parity check equations:
\begin{equation} \label{eq:vv}
\sum_{i=1}^n \sum_{j=1}^s \lambda_{i,j,a_i}^t c_{i,j,\underline{a}} = 0, \quad\quad  t=0,1,\dots,(n-k)s-1,
 \quad \underline{a}\in\{0,1,\dots,d-k\}^n.
\end{equation}
We can see that for each fixed $\underline{a}\in\{0,1,\dots,d-k\}^n$, the vector 
$(c_{i,j,\underline{a}}:i\in[n],j\in[s])$ forms a Generalized Reed-Solomon (GRS) code with length $sn$ and dimension $sk$,
so $\cC$ is indeed an $(n,k,l=s(d-k+1)^n)$ MDS array code.

\begin{proposition}
The code $\cC$ has optimal $\alpha$-decoding radius.
\end{proposition}
\begin{proof}
From each $C_i$ we download $f(C_i):=(c_{i,j,\underline{a}}:j\in[m],\underline{a}\in\{0,1,\dots,d-k\}^n) \in F^{m(d-k+1)^n}$, which contains a $m/s=\alpha$ proportion of coordinates in $C_i$.
Since $(c_{i,j,\underline{a}}:i\in[n],j\in[s])$ forms an $(sn,sk)$ MDS code for every $\underline{a}\in\{0,1,\dots,d-k\}^n$, we can calculate $(c_{i,j,\underline{a}}:i\in[n],j\in[s])$ from $\{f(C_i):i\in\cI\}$ for every $\underline{a}\in\{0,1,\dots,d-k\}^n$ and every subset $\cI\subseteq[n]$ with cardinality $|\cI|\ge sk/m=k/\alpha$.
In other words, we can recover the original codeword $(C_1,C_2,\dots,C_n)$ from $\{f(C_i):i\in\cI\}$ from every subset $\cI\subseteq[n]$ with cardinality $|\cI|\ge k/\alpha$.
We thus conclude that we can do fractional decoding up to $\lfloor(n-k/\alpha)/2\rfloor$ errors.
\end{proof}

\begin{proposition}
The code $\cC$ has the $d$-optimal repair property.
\end{proposition}
\begin{proof}
Without loss of generality suppose that we want to repair $C_1$. For $u\in\{0,1,\dots,d-k\}$, we write 
$\underline{a}(1,u):=(u,a_2,a_3,\dots,a_n)$, namely we replace $a_1$ with $u$ in vector $\underline{a}$ to obtain $\underline{a}(1,u)$. 
Replacing $\underline{a}$ with $\underline{a}(1,u)$ in \eqref{eq:vv}, we obtain that for every $u\in\{0,1,\dots,d-k\}$,
$$
\sum_{j=1}^s \lambda_{1,j,u}^t c_{1,j,\underline{a}(1,u)} +
\sum_{i=2}^n \sum_{j=1}^s \lambda_{i,j,a_i}^t c_{i,j,\underline{a}(1,u)} = 0, \quad\quad  t=0,1,\dots,(n-k)s-1,
 \quad \underline{a}\in\{0,1,\dots,d-k\}^n.
$$
Summing these equations over $u\in\{0,1,\dots,d-k\}$, we have
\begin{align*}
\sum_{u=0}^{d-k}\sum_{j=1}^s \lambda_{1,j,u}^t c_{1,j,\underline{a}(1,u)} +
\sum_{i=2}^n \sum_{j=1}^s  &  \lambda_{i,j,a_i}^t \Big(\sum_{u=0}^{d-k} c_{i,j,\underline{a}(1,u)}\Big) = 0, \\
 &  t=0,1,\dots,(n-k)s-1,
 \quad \underline{a}\in\{0,1,\dots,d-k\}^n.
\end{align*}
Since all the $\lambda$'s in the equation above are distinct, we conclude that for every fixed $\underline{a}\in\{0,1,\dots,d-k\}^n$, the vector 
\begin{equation}\label{eq:ww}
\Big(\{c_{1,j,\underline{a}(1,u)}:u\in\{0,1,\dots,d-k\},j\in[s]\},
\Big\{\sum_{u=0}^{d-k} c_{i,j,\underline{a}(1,u)}:i\in\{2,3,\dots,n\},j\in[s] \Big\}\Big)
\end{equation}
 forms a GRS code with length $s(d-k+1)+s(n-1)=s(d-k+n)$ and dimension $s(d-k+n)-s(n-k)=sd$.
As an immediate consequence, we can calculate the vector in \eqref{eq:ww} from 
$$
\Big\{\sum_{u=0}^{d-k} c_{i,j,\underline{a}(1,u)}:i\in\cR,j\in[s] \Big\}
$$
for any subset $\cR\subseteq[n]$ with cardinality $|\cR|=d$.
Therefore we can download the following 
$\frac{dl}{d-k+1}$ symbols in $F$
$$
\Big\{\sum_{u=0}^{d-k} c_{i,j,\underline{a}(1,u)}:i\in\cR,j\in[s],\underline{a}\in\{0,1,\dots,d-k\}^n,a_1=0 \Big\}
$$
from the $d$ helper nodes $\{C_i:i\in\cR\}$, and we will be able to calculate
\begin{align*}
& \{c_{1,j,\underline{a}(1,u)}:u\in\{0,1,\dots,d-k\},j\in[s],\underline{a}\in\{0,1,\dots,d-k\}^n,a_1=0\} \\
= & \{c_{1,j,\underline{a}},j\in[s],\underline{a}\in\{0,1,\dots,d-k\}^n\},
\end{align*}
which is the set of all the coordinates of $C_1$.
This completes the proof of the $d$-optimal repair property.
\end{proof}

\bibliographystyle{IEEEtran}
\bibliography{fractional}

\end{document}